\documentclass[10pt,a4paper,onecolumn]{article}
\usepackage{marginnote}
\usepackage{graphicx}
\usepackage{orcidlink}
\usepackage{xcolor}
\usepackage{authblk,etoolbox}
\usepackage{titlesec}
\usepackage{calc}
\usepackage{tikz}
\usepackage{fontawesome}
\usepackage{hyperref}
\usepackage{microtype}
\hypersetup{colorlinks,breaklinks,
            urlcolor=[rgb]{0.0, 0.5, 1.0},
            linkcolor=[rgb]{0.0, 0.5, 1.0}}
\usepackage{caption}
\usepackage{tcolorbox}
\usepackage{amssymb,amsmath}
\usepackage{ifxetex,ifluatex}
\usepackage{seqsplit}
\usepackage{xstring}

\usepackage{fixltx2e} 
\usepackage[
  backend=biber,
  citestyle=authoryear
]{biblatex}
\bibliography{paper.bib}

\let\textttOrig=\texttt
\def\texttt#1{\expandafter\textttOrig{\seqsplit{#1}}}
\renewcommand{\seqinsert}{\ifmmode
  \allowbreak
  \else\penalty6000\hspace{0pt plus 0.02em}\fi}

\makeatletter
\let\href@Orig=\href
\def\href@Urllike#1#2{\href@Orig{#1}{\begingroup
    \def\Url@String{#2}\Url@FormatString
    \endgroup}}
\def\href@Notdoi#1#2{\def\tempa{#1}\def\tempb{#2}%
  \ifx\tempa\tempb\relax\href@Urllike{#1}{#2}\else
  \href@Orig{#1}{#2}\fi}
\def\href#1#2{%
  \IfBeginWith{#1}{https://doi.org}%
  {\href@Urllike{#1}{#2}}{\href@Notdoi{#1}{#2}}}
\makeatother

\usepackage[top=3.2cm, bottom=2.9cm, right=1.5cm, left=1.0cm,
            headheight=1.9cm, reversemp, includemp, marginparwidth=3.9cm]{geometry}

\titleformat{\section}
  {\normalfont\sffamily\Large\bfseries}
  {}{0pt}{}
\titleformat{\subsection}
  {\normalfont\sffamily\large\bfseries}
  {}{0pt}{}
\titleformat{\subsubsection}
  {\normalfont\sffamily\bfseries}
  {}{0pt}{}
\titleformat*{\paragraph}
  {\sffamily\normalsize}

\usepackage{fancyhdr}
\makeatletter
\let\ps@plain\ps@fancy
\fancyheadoffset[L]{4.5cm}
\fancyfootoffset[L]{4.5cm}

\definecolor{linky}{rgb}{0.0, 0.5, 1.0}

\newtcolorbox{repobox}
   {colback=red, colframe=red!75!black,
     boxrule=0.5pt, arc=2pt, left=6pt, right=6pt, top=3pt, bottom=3pt}

\newcommand{\ExternalLink}{%
   \tikz[x=1.2ex, y=1.2ex, baseline=-0.05ex]{%
       \begin{scope}[x=1ex, y=1ex]
           \clip (-0.1,-0.1)
               --++ (-0, 1.2)
               --++ (0.6, 0)
               --++ (0, -0.6)
               --++ (0.6, 0)
               --++ (0, -1);
           \path[draw,
               line width = 0.5,
               rounded corners=0.5]
               (0,0) rectangle (1,1);
       \end{scope}
       \path[draw, line width = 0.5] (0.5, 0.5)
           -- (1, 1);
       \path[draw, line width = 0.5] (0.6, 1)
           -- (1, 1) -- (1, 0.6);
       }
   }

\patchcmd{\@maketitle}{center}{flushleft}{}{}
\patchcmd{\@maketitle}{center}{flushleft}{}{}
\patchcmd{\@maketitle}{\LARGE}{\LARGE\sffamily}{}{}
\def\maketitle{{%
  
  \AB@maketitle}}
\makeatletter
\renewcommand\AB@affilsepx{ \protect\Affilfont}
\renewcommand\AB@affilnote[1]{{\bfseries #1}\hspace{3pt}}
\renewcommand{\affil}[2][]%
   {\newaffiltrue\let\AB@blk@and\AB@pand
      \if\relax#1\relax\def\AB@note{\AB@thenote}\else\def\AB@note{#1}%
        \setcounter{Maxaffil}{0}\fi
        \begingroup
        \let\href=\href@Orig
        \let\texttt=\textttOrig
        \let\protect\@unexpandable@protect
        \def\thanks{\protect\thanks}\def\footnote{\protect\footnote}%
        \@temptokena=\expandafter{\AB@authors}%
        {\def\\{\protect\\\protect\Affilfont}\xdef\AB@temp{#2}}%
         \xdef\AB@authors{\the\@temptokena\AB@las\AB@au@str
         \protect\\[\affilsep]\protect\Affilfont\AB@temp}%
         \gdef\AB@las{}\gdef\AB@au@str{}%
        {\def\\{, \ignorespaces}\xdef\AB@temp{#2}}%
        \@temptokena=\expandafter{\AB@affillist}%
        \xdef\AB@affillist{\the\@temptokena \AB@affilsep
          \AB@affilnote{\AB@note}\protect\Affilfont\AB@temp}%
      \endgroup
       \let\AB@affilsep\AB@affilsepx
}
\makeatother

\renewcommand\Affilfont{\sffamily\small\mdseries}
\ifnum 0\ifxetex 1\fi\ifluatex 1\fi=0 
  \usepackage[T1]{fontenc}
  \usepackage[utf8]{inputenc}

\else 
  \ifxetex
    \usepackage{mathspec}
  \else
    \usepackage{fontspec}
  \fi
  \defaultfontfeatures{Ligatures=TeX,Scale=MatchLowercase}

\fi
\IfFileExists{upquote.sty}{\usepackage{upquote}}{}
\IfFileExists{microtype.sty}{%
\usepackage{microtype}
\UseMicrotypeSet[protrusion]{basicmath} 
}{}

\usepackage{hyperref}
\hypersetup{unicode=true,
            pdftitle={\texttt{WDM\_GW\_Wavelets}: A fast, \texttt{JAX}-based Python implementation of the Wilson-Daubechies-Meyer wavelet transform for the time-frequency analysis of gravitational wave data},
            pdfauthor={Moore},
            pdfborder={0 0 0},
            breaklinks=true}
\let\addcontentslineOrig=\addcontentsline
\def\addcontentsline#1#2#3{\bgroup
  \let\texttt=\textttOrig\addcontentslineOrig{#1}{#2}{#3}\egroup}
\let\markbothOrig\markboth
\def\markboth#1#2{\bgroup
  \let\texttt=\textttOrig\markbothOrig{#1}{#2}\egroup}
\let\markrightOrig\markright
\def\markright#1{\bgroup
  \let\texttt=\textttOrig\markrightOrig{#1}\egroup}

\usepackage{graphicx,grffile}
\makeatletter
\def\maxwidth{\ifdim\Gin@nat@width>\linewidth\linewidth\else\Gin@nat@width\fi}
\def\maxheight{\ifdim\Gin@nat@height>\textheight\textheight\else\Gin@nat@height\fi}
\makeatother
\setkeys{Gin}{width=\maxwidth,height=\maxheight,keepaspectratio}
\IfFileExists{parskip.sty}{%
\usepackage{parskip}
}{
\setlength{\parindent}{0pt}
\setlength{\parskip}{6pt plus 2pt minus 1pt}
}
\ifx\paragraph\undefined\else
\let\oldparagraph\paragraph
\renewcommand{\paragraph}[1]{\oldparagraph{#1}\mbox{}}
\fi
\ifx\subparagraph\undefined\else
\let\oldsubparagraph\subparagraph
\renewcommand{\subparagraph}[1]{\oldsubparagraph{#1}\mbox{}}
\fi

\title{\texttt{WDM\_GW\_Wavelets}: A fast, \texttt{JAX}-based Python implementation of the Wilson-Daubechies-Meyer wavelet transform for the time-frequency analysis of gravitational wave data}

        \author[1,2,3]{Christopher J. Moore\,\orcidlink{0000-0002-2527-0213}}

        \author[1,2]{Tomasz Kinowski\,\orcidlink{0009-0004-9474-7654}}
    
      \affil[1]{Institute of Astronomy, University of Cambridge, Madingley Road, Cambridge, CB3 0HA, UK}
      
      \affil[2]{Kavli Institute for Cosmology, University of Cambridge, Madingley Road, Cambridge, CB3 0HA, UK}
      
      \affil[3]{Department of Applied Mathematics and Theoretical Physics, Centre for Mathematical Sciences, University of Cambridge, Wilberforce Road, Cambridge, CB3 0WA, UK}

  \date{\vspace{-5ex}}

\begin{document}
\maketitle

\marginpar{
  \sffamily\small

  {\bfseries DOI:} \href{https://doi.org/10.5281/zenodo.21890468}{\color{linky}{10.5281/zenodo.21890468}}

  \vspace{2mm}

  {\bfseries Software}
  \begin{itemize}
    \setlength\itemsep{0em}
    \item \href{https://github.com/}{\color{linky}{Review}} \ExternalLink
    \item \href{https://github.com/cjm96/WDM_GW_wavelets}{\color{linky}{Repository}} 
    \ExternalLink
    \item \href{https://doi.org/10.5281/zenodo.21890468}{\color{linky}{Archive}} \ExternalLink
  \end{itemize}

  \vspace{2mm}

  {\bfseries Submitted:} 11 August 2026\\
  {\bfseries Published:} NA

}

\vspace{-2mm}
\hypertarget{summary}{%
\section{Summary}\label{summary}}
\vspace{-2mm}

Transient gravitational-wave (GW) signals observed by current ground-based interferometers are short; mostly $\sim 0.5-30\,\mathrm{s}$. 
These signals are usually conveniently analysed in the frequency domain where the instrumental noise (usually assumed to be stationary) is easily modelled.
The next generation of ground-based interferometers, as well as the upcoming LISA space-based interferometer, will observe much longer signals; several hours for ground-based instruments and years for space-based ones.
The instrumental noise will not remain stationary over these time scales.
Additionally, the detectors will move and rotate appreciably over these time scales, complicating the modelling of the instrument response.

\begin{figure}[t]
    \centering
    \includegraphics[width=0.65\linewidth]{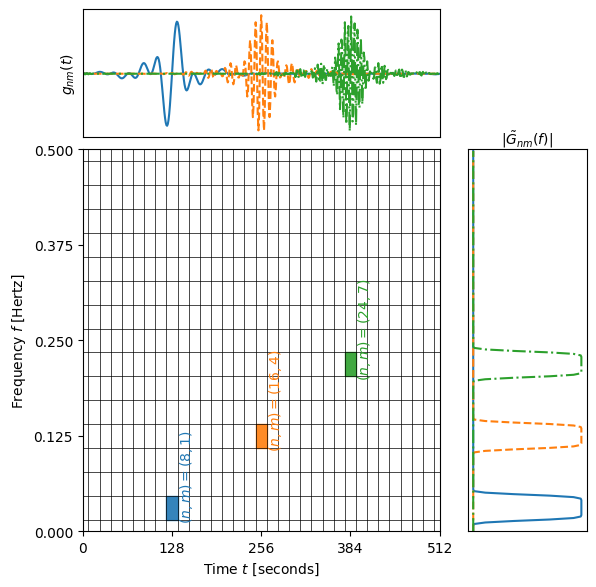}
    \caption{ \label{fig:time_freq_plane}
        Example WDM wavelets plotted in both time (top) and frequency (right) domains. The  time-frequency grid shows where the wavelets have support. Figure reproduced from the package documentation.
    }
\end{figure}

These challenges motivate the development of time-frequency methods for GW data analysis.
The Wilson-Daubechies-Meyer (WDM) wavelet basis introduced by \cite{Necula:2012}, and previously used in a GW context by Coherent WaveBurst (\texttt{cWB}) (see, for example, \cite{Klimenko:2016} and \cite{Drago:2021}), has been identified as a promising set of basis functions for this purpose by \cite{Cornish:2020}. 
The WDM wavelets expand a time series signal $x(t)$ of length $N$ using a discrete, orthogonal basis of wave packets,
\begin{equation}
    x(t) = \sum_{n=0}^{N_t-1}\; \sum_{m=0}^{N_f-1} \, w_{nm} \, g_{nm}(t) \, ,
\end{equation}
where $N=N_tN_f$ and the basis functions $g_{nm}$ are localised in time around $t_n=n\Delta T$ and have compact support in frequency around $f_m=m\Delta F$. 
The quantities $\Delta T$ and $\Delta F$ are the wavelet time and frequency resolutions respectively; 
one may be chosen freely while the other is constrained by $\Delta T \Delta F = 1/2$.
The WDM basis maps the time series onto a regular grid of pixels that cover the time-frequency plane uniformly (see figure \ref{fig:time_freq_plane}).
The discrete WDM wavelet transform can be written in a form that makes its computational cost comparable to that of a fast Fourier transform.

GW data analysis often requires introducing time delays between different detectors or data streams, either to synchronize them with respect to a particular sky location or to eliminate a common noise source. 
A key property of the WDM wavelets is the existence of analytic time-delay filters which allow time delays to be performed in the time-frequency domain. 

\vspace{-2mm}
\hypertarget{statement}{%
\section{Statement of need}\label{statement}}
\vspace{-2mm}

There are many different methods for time series data analysis in the time-frequency domain.
Similarly, there are many families of wavelets and different conventions within each family.
This can lead to difficulties when comparing results between codes and authors.
There is a need in the GW data analysis community for a single convention with a stable, publicly available implementation that is well-documented and has the mathematical details and conventions clearly described alongside the code.
For maximum utility, this should include the time-delay filters along with the transform itself.
Additionally, to allow for Bayesian data analysis to be performed in the time-frequency domain, there is a need for the wavelet transform to be fast and to be executable on multiple backends, including CPUs, GPUs, and TPUs.

\texttt{WDM\_GW\_Wavelets} is designed to meet these needs.
\texttt{WDM\_GW\_Wavelets} is a \texttt{JAX}-compatible Python package that implements the WDM wavelet transform and the associated time-delay filters and is accompanied by mathematical documentation describing the conventions used.

\vspace{-2mm}
\hypertarget{state}{%
\section{State of the field}\label{state}}
\vspace{-2mm}

The WDM wavelets have been used previously in the context of GW data analysis (see, for example, \cite{Necula:2012}, \cite{Klimenko:2016} and \cite{Drago:2021}).
However, \cite{Cornish:2020} was the first to advocate the benefits of this particular wavelet family for performing GW data analysis in the LISA context. 
\cite{Cornish:2020} gives only brief mathematical details of the WDM wavelets (for example, it does not discuss the special handling needed for the $m=0$ coefficients which store the zero and Nyquist frequency components of the signal) and directs the reader to \cite{Necula:2012} for these details.
\cite{Cornish:2020} includes C code \href{https://github.com/eXtremeGravityInstitute/WDM_Transform}{(\faGithub)} for performing the wavelet transform.

Multiple groups are actively investigating the use of WDM wavelets for GW data analysis.
While this \texttt{WDM\_GW\_Wavelets} package was being developed, the preprints \cite{Johnson:2026} and \cite{Vajpeyi:2026} appeared. 
\cite{Vajpeyi:2026} is accompanied by a public package \href{https://github.com/pywavelet/wdm_transform}{(\faGithub)} that implements the WDM transform (but currently not the filters).

\texttt{WDM\_GW\_Wavelets} provides a well-documented and stable reference implementation that combines both the WDM transform and the time-delay filters with \texttt{JAX} compatibility in a single package.
\texttt{WDM\_GW\_Wavelets} also enables the use of Bayesian inference algorithms that require access to derivatives through \texttt{JAX's} automatic differentiation (autodiff) functionality.

\vspace{-2mm}
\hypertarget{software}{%
\section{Software design}\label{software}}
\vspace{-2mm}

\texttt{WDM\_GW\_Wavelets}'s design is kept simple with the main functionality contained in a single class and with the methods aligned directly with the mathematical description in the \href{https://cjm96.github.io/WDM_GW_wavelets/}{documentation}.
This was done to make the definitions and conventions of the wavelets used as clear as possible and to allow for comparison with other authors and codes.
\texttt{JAX} is used throughout as this allows the code to be run on different backends (e.g.\ CPU, GPU or TPU) as well as allowing for JIT compilation and automatic differentiation to be used downstream in Bayesian inference.

\vspace{-2mm}
\hypertarget{research}{%
\section{Research impact statement}\label{research}}
\vspace{-2mm}

\texttt{WDM\_GW\_Wavelets} is being actively used in the context of the UK LISA Ground Segment for the development of tools to address data analysis challenges for LISA, including the global fit. The first publication by the authors to use this package focuses on modelling the LISA response function to GWs (\cite{Kinowski:2026} and \href{https://github.com/tomaszkinowski/WDM_LISAresponse}{\faGithub}).
\texttt{WDM\_GW\_Wavelets} has also been used to make comparisons with the results of other authors and codes (\cite{Vajpeyi:2026} and \href{https://github.com/pywavelet/wdm_transform}{\faGithub}).

\vspace{-2mm}
\hypertarget{ai}{%
\section{AI usage disclosure}\label{ai}}
\vspace{-2mm}

Parts of the \texttt{WDM\_GW\_Wavelets} codebase, particularly those associated with performance and JIT-compilation with \texttt{JAX}, were created with the assistance of the generative AI tool \texttt{Claude}. 
However, no code was generated wholesale by AI, and all code has been verified by human developers. 
Other than spelling and grammar checking, this paper was written without AI.

\vspace{-2mm}
\hypertarget{acknowledgements}{%
\section{Acknowledgements}\label{acknowledgements}}
\vspace{-2mm}

We gratefully acknowledge many useful discussions on the topic of GW data 
analysis in the time-frequency domain with members of the UK LISA Ground 
Segment team:
Alberto Vecchio, Hannah Middleton, Geraint Pratten, Christian Chapman-Bird,
Ian Harry, Michael Williams, 
Graham Woan, Christopher Berry, Alexander (Ollie) Burke, and 
Leor Barack.

\vspace{-2mm}
\hypertarget{references}{%
\section*{References}\label{references}}
\addcontentsline{toc}{section}{References}
\vspace{-2mm}

\leavevmode\hypertarget{ref-Necula:2012}{}%
Necula, Klimenko, and Mitselmakher 2012. ``Transient analysis with fast Wilson-Daubechies time-frequency transform'' \emph{Journal of Physics Conference Series} 363, 012032. \url{https://doi.org/10.1088/1742-6596/363/1/012032}.

\leavevmode\hypertarget{ref-Klimenko:2016}{}%
Klimenko \emph{et al.} 2016. ``Method for detection and reconstruction of gravitational wave transients with networks of advanced detectors'' \emph{Physical Review D} 93, 4, 042004. \url{https://doi.org/10.1103/PhysRevD.93.042004}.

\leavevmode\hypertarget{ref-Drago:2021}{}%
Drago \emph{et al.} 2021. ``coherent WaveBurst, a pipeline for unmodeled gravitational-wave data analysis'' \emph{SoftwareX} 14, 100678. \url{https://doi.org/10.1016/j.softx.2021.100678}.

\leavevmode\hypertarget{ref-Cornish:2020}{}%
Cornish 2020. ``Time-frequency analysis of gravitational wave data'' \emph{Physical Review D} 102, 12, 124038. \url{https://doi.org/10.1103/PhysRevD.102.124038}.

\leavevmode\hypertarget{ref-Johnson:2026}{}%
Johnson, Chatziioannou and Summers 2026. ``The WDM Time-Frequency Transform in Gravitational-Wave Data Analysis I: Formalism'' \url{https://doi.org/10.48550/arXiv.2606.21473}.

\leavevmode\hypertarget{ref-Vajpeyi:2026}{}%
Vajpeyi \emph{et al.} 2026. ``An explicit and differentiable Wilson-Daubechies-Meyer transform for gravitational-wave data analysis'' \url{https://doi.org/10.48550/arXiv.2606.20269}.

\leavevmode\hypertarget{ref-Kinowski:2026}{}%
Kinowski and Moore 2026. ``Modeling the LISA response in the time-frequency domain'' (in preparation).

\end{document}